\documentclass{elsart} 
\usepackage{graphicx,amssymb,amsmath}

\begin{document}

\begin{frontmatter}

\title{Ubiquity of metastable-to-stable crossover in weakly
  chaotic dynamical systems}

\author[rio]{Fulvio Baldovin}, \ead{baldovin@cbpf.br}
\author[rio]{Luis G. Moyano}, \ead{moyano@cbpf.br}
\author[cordoba]{Ana P. Majtey}, \ead{amajtey@famaf.unc.edu.ar}
\author[mexico]{Alberto Robledo} and
\ead{robledo@fisica.unam.mx} 
\author[rio]{Constantino Tsallis} \ead{tsallis@cbpf.br}

\address[rio]{Centro Brasileiro de Pesquisas F\'{\i}sicas,\\
Rua Xavier Sigaud 150, 22290-180 Rio de Janeiro -- RJ,
Brazil } \address[cordoba]{Facultad de Matem\'{a}tica,
Astronom\'{\i}a y F\'{\i}sica, Universidad Nacional de
C\'{o}rdoba, Ciudad Universitaria, 5000, C\'{o}rdoba,
Argentina}

\address[mexico]{Instituto de F\'{\i}sica, Universidad
Nacional Aut\'{o}noma de M\'{e}xico,\\ Apartado Postal
20-364, M\'{e}xico 01000 D.F., Mexico}
%\date{July 15, 2002}

\begin{abstract}
We present a comparative study of several dynamical systems
of increasing 
complexity, namely, the logistic map with additive noise,
one, two and many globally-coupled 
standard maps, and the Hamiltonian Mean Field model
(i.e., the classical inertial infinitely-ranged
ferromagnetically coupled XY spin model). 
We emphasize the appearance, in all of these systems, of
metastable states and 
their ultimate crossover to the equilibrium state. We comment on the
underlying mechanisms responsible for these phenomena (weak
chaos) and compare common 
characteristics. 
We point out that this ubiquitous behavior appears to be
associated to the features of the nonextensive generalization of the
Boltzmann-Gibbs statistical mechanics.
\end{abstract}


\end{frontmatter}

%Alternative title:

%\textbf{Ubiquity of metastable to stable crossover in weakly
%chaotic dynamical systems and dynamically-stalled
%long-ranged interacting systems}

\section{Introduction}
The statistical-mechanical comprehension of the occurrence
and eventual disappearance of out-of-equilibrium
quasistationary states (QSS), referred here also as
metastable states, is an important undertaking that nowadays
attracts the attention of researchers in various fields,
ranging from weakly chaotic nonlinear dynamics 
to slow glassy dynamics in condensed matter. 
Under specific classes of initial conditions,
the dynamics in these systems displays two-step relaxation
processes, entering first one or more periods of little variation
before a crossover to the final equilibrium state. The QSS
are conjectured to emerge as the result of long-ranged
correlations or interactions amongst the elements of the
system or long-ranged time correlations in the orbits of
iterated maps. An interesting statistical-mechanical model
that displays QSS is the Hamiltonian Mean Field (HMF) model
\cite{antoni_01}, that in its simplest version consists
of a set of $N$ inertial XY classical spins or rotors all
of which interact equally with each other. But also,
interestingly, the most relevant features of QSS dynamics
are displayed by considerably simpler nonlinear dynamical
maps, as it is the case of conservative coupled maps
\cite{baldovin_01} and also by single, dissipative,
one-dimensional maps, like the prototypical logistic map
close to the edge of chaos and in the presence of external
noise \cite{robledo_01}.

In this paper we present a comparative study of the
aforementioned models, starting our discussion with
low-dimensional nonlinear maps and ending it with the HMF
model. We emphasize the analogies found between their
dynamical properties and indicate the mechanisms that give
raise to the QSS as well as those responsible for the
crossover to the equilibrium state. 
In all of these cases the quantities that describe dynamical
evolution seem to be associated to the predictions of the
nonextensive generalization \cite{tsallis_01} of the 
Boltzmann-Gibbs (BG) statistical mechanics.
Also, in all of
these examples there is clear evidence of the
non-commutability of the infinite time limit with
other important parameter limits, such as the thermodynamic limit
($N\rightarrow \infty $) or the limit of vanishing
noise amplitude. 
Moreover, there is a common indication that, in the
characterization of the phase space available for dynamical
evolution, there appears evidence of fractalization and of
ergodicity failure. 

The paper is organized as follows. In the next section we
describe how the dynamics at the edge of chaos of the
logistic map leads, via the addition of external noise, to the
crossover from QSSs to a strongly chaotic state that is the
analog of the BG equilibrium state \cite{robledo_01}. We
present here new numerical evidence by means of an ensemble
implementation of the dynamics in this system. In section
\ref{section_cons_maps} we note that the emergence of QSSs
and eventual departure from them appear associated to the
complex structure of phase space accessibility in symplectic
maps governed by the Kolmogorov-Arnold-Moser (KAM)
theorem \cite{baldovin_01}. 
We also show new results that indicate that the
QSS, present at the lowest possible dimensions, 
persists as hundreds of symplectic maps are
(globally) coupled. Section \ref{section_HMF} is devoted to
the QSS of the HMF model.  Particular attention is paid to
the time evolution of the anomalous probability distribution
functions (PDFs) for the velocities, and also to basic
questions referring to the measurability of temperature and
to the zeroth principle of thermodynamics as it relates to
the QSS \cite{moyano_01}. Finally, in section
\ref{section_discussion} we draw our conclusions, we make
reference to the existing analogies amongst the different
models described and discuss the underlying mechanism that
gives raise to the QSS.

\section{QSS in the logistic map with external noise}
\label{section_logistic} 
Let us start our analysis by
recalling \cite{robledo_01} that phenomena like two-step
relaxation and aging can be displayed at the lowest possible
dimension, namely, by means of a one-dimensional map. For
this purpose we consider the paradigmatic logistic map with
additive white noise,
\begin{equation}
x_{\tau+1}=1-\mu x_{\tau}^{2}+\xi _{\tau}\sigma
,\;\;\;\;\;\;\tau=0,1,...,\;\;\;x\in [-1,1],\;\;\;\mu \in
[0,2], \label{eq_logistic}
\end{equation}
where $\xi _{\tau}$ is Gaussian-distributed and
delta-correlated,  $\langle \xi _{\tau}\rangle
_{\xi }=0$,     
\mbox{$\langle \xi _{\tau}\xi _{\tau^{\prime }}\rangle
_{\xi }=\delta _{\tau\tau^{\prime }}$}, 
and $\sigma $ measures the
noise intensity. As it is well known, for $\sigma =0$ the
Feigenbaum attractor at the edge of chaos $\mu =\mu
_{c}\equiv 1.401155189...  $ is the accumulation point of
both the period-doubling and band-splitting cascades. The
Lyapunov coefficient is zero \cite{grassberger_01}
and the attractor is a cantor
set of fractal dimension $d_{f}=0.5338...$. It has been
analytically proved by means of the conventional
renormalization-group approach for this type of map that the
dynamics at this critical point is unmistakably associated
to the nonextensive statistical mechanics
\cite{baldovin_02}. Specifically, the sensitivity to initial
conditions is actually given in a closed form by a
$q$-exponential function and it is related to the rate of
entropy production via a $q$-generalized Pesin identity,
linking the sensitivity to initial conditions to the
nonextensive expression for the entropy, $S_{q}=\ln _{q}W$
(where $W$ is the number of equally-probable positions of an
ensemble of iterates at a given time). For this stationary
state the index $q$ takes the specific value
$q=0.2445...$. Let us remind the reader that the
$q$-exponential and its inverse the $q$-logarithm are
defined, respectively, as $\exp _{q}(x)\equiv
[1+(1-q)x]^{1/1-q}$ and $\ln _{q}(x)\equiv
(x^{1-q}-1)/(1-q)$, and that these expressions reduce to the
usual exponential and logarithmic functions in the limit
$q\to 1$.

\begin{figure}
\begin{center}
\includegraphics[width=14cm,angle=0]{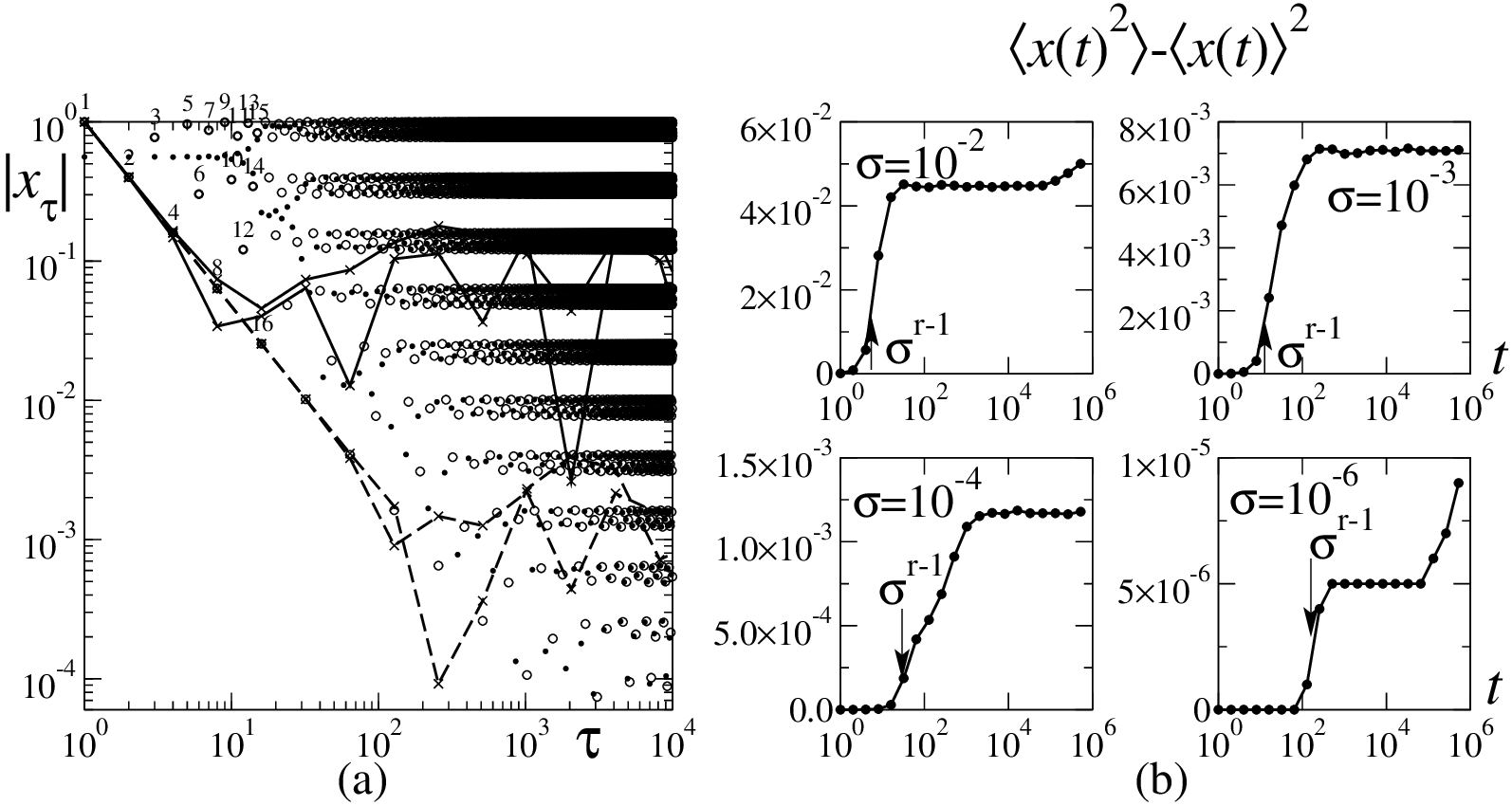}
\end{center}
\caption{\small 
(a) Phase space structure of the logistic map of the logistic map 
with and without additive noise. 
Empty circles correspond to the attractor at $\mu=\mu_c$ and  
$\sigma=0$, obtained by iterating $x_{0}=0$ 
(the numbers label time $\tau =1,...,16$). Small dots
correspond to $x_{0}=0.56023$, 
close to a repeller, the unstable solution of 
$x=1-\mu _{c}x^{2}$.
Full (and dashed) lines represent two trajectories for
$\mu=\mu_c$, $\sigma=10^{-3}$ (and $\sigma=10^{-6}$).
(b) Plateaux structure. Time evolution 
(using the shifted time $t=2^n-1$)
of the variance
of an ensemble of $M=10^4$ trajectories all starting at $x_0=0$,
for different orders of magnitude of the noise strength
$\sigma$. 
Notice that the crossover times coincide with $\sigma^{r-1}$
(with $r\simeq 0.633$)
indicated with arrows (see text). 
}
\label{fig_logistic_1}
\end{figure}
More precisely, the Feigenbaum attractor can be described by
means of a set of position subsequences generated when the
iterative map is given the initial position $x_{0}=0$
\cite{baldovin_02}. See the empty circles in Fig.
\ref{fig_logistic_1}(a) where we have plotted the absolute
value of these positions $\left| x_{\tau }\right| $ and time
$\tau $ at which they are reached in logarithmic
scales. Each of the subsequences is labelled by an integer
$k=0,1,...$ and each is generated by the time subsequence
$\tau _{k}=(2k+1)2^{n-k}$, with $n=0,1,...$ and 
$n\geq k$. This can be appreciated in
Fig. \ref{fig_logistic_1}(a). In the following we will focus
on the principal subsequence $\tau _{0}=2^{n}$. Another
important feature of Fig.  \ref{fig_logistic_1}(a) is that the
logarithm of the attractor positions appear clearly divided
into an infinite number of equally-spaced horizontal
bands. In the empty space between any two bands there are
points called \textit{repellers}, the unstable solutions of
the iterated equation $x=1-\mu _{c}x^{2}$. (See the small
dots in Fig. \ref{fig_logistic_1}(a)). At the chaos threshold
of the logistic map with $\sigma \neq 0$ the dynamics is
governed by two competing phenomena: On the one hand the
action of the noise is to drive the iterates out of the
bands, and on the other hand the instability in the
neighborhood of the repellers sends the positions back into
the attractor bands. When the noise intensity is large
enough to fill a gap between bands, the result is that the
iterates are able to jump from one band to the next. (See
full and dashed lines in Fig. \ref{fig_logistic_1}(a)).  Since
the band gaps in $\left| x\right| $-space are of variable
size, increasing in size as $\left| x\right| \rightarrow 1$,
the motion becomes confined among some few bands for a
certain amount of time but at longer times the iterates are
able to spread to other more distant bands.

The analysis of the crossover time $t_c$ for the
broadening of the attractor bands produced by the noise can
be made quantitative. At the edge of chaos $\mu =\mu _{c}$,
for $\sigma \neq 0$, the iterated position of $x_{0}=0$ at
shifted time $t=\tau _{0}-1$ is given by (see
\cite{robledo_01} for details):
\begin{equation}
x_{t}=\exp _{2-q}(-\lambda _{q}t)[1+\xi _{t}\sigma \exp
_{r}(\lambda _{r}t)],
\end{equation}
where $q=1-\ln 2/\ln \alpha =0.2445...$, $\lambda _{q}=\ln
\alpha /\ln 2=1.3236...$ is the generalized Lyapunov
coefficient, $r=1-\ln 2/\ln \kappa \simeq 0.633$, $\lambda
_{r}=\ln \kappa /\ln 2\simeq 2.727$ ($\alpha =2.50290...$
being one of Feigenbaum's universal constants and $\kappa
\simeq 6.619$ a constant associated to the noise term). Let
us consider now a different trajectory $x_{t}^{\prime }$
also starting at the origin at $t=0$, $x_{0}^{\prime
}=0$. Since the noise acts independently on each trajectory,
after time $t$ the orbit's separation is given by
\begin{equation}
x_{t}-x_{t}^{\prime }=\exp _{2-q}(-\lambda _{q}t)\exp
_{r}(\lambda _{r}t)\sigma (\xi _{t}-\xi _{t}^{\prime }).
\end{equation}
At time $t=2^{n}-1$ the separation between bands at which
the iterate is located is of order $\exp _{2-q}(-\lambda
_{q}t)$. Using \mbox{$\xi_t-\xi'_t\sim 1$} we obtain an
estimate of the crossover time $t_c$ at which the
noise-assisted jumps into different bands starts to take
place. This is
\begin{equation}
\exp _{r}(\lambda _{r}t_{c})\sigma \sim 1\Rightarrow
t_c\sim \sigma ^{r-1}.
\end{equation}

\begin{figure}
\begin{center}
\includegraphics[width=10cm,angle=0]{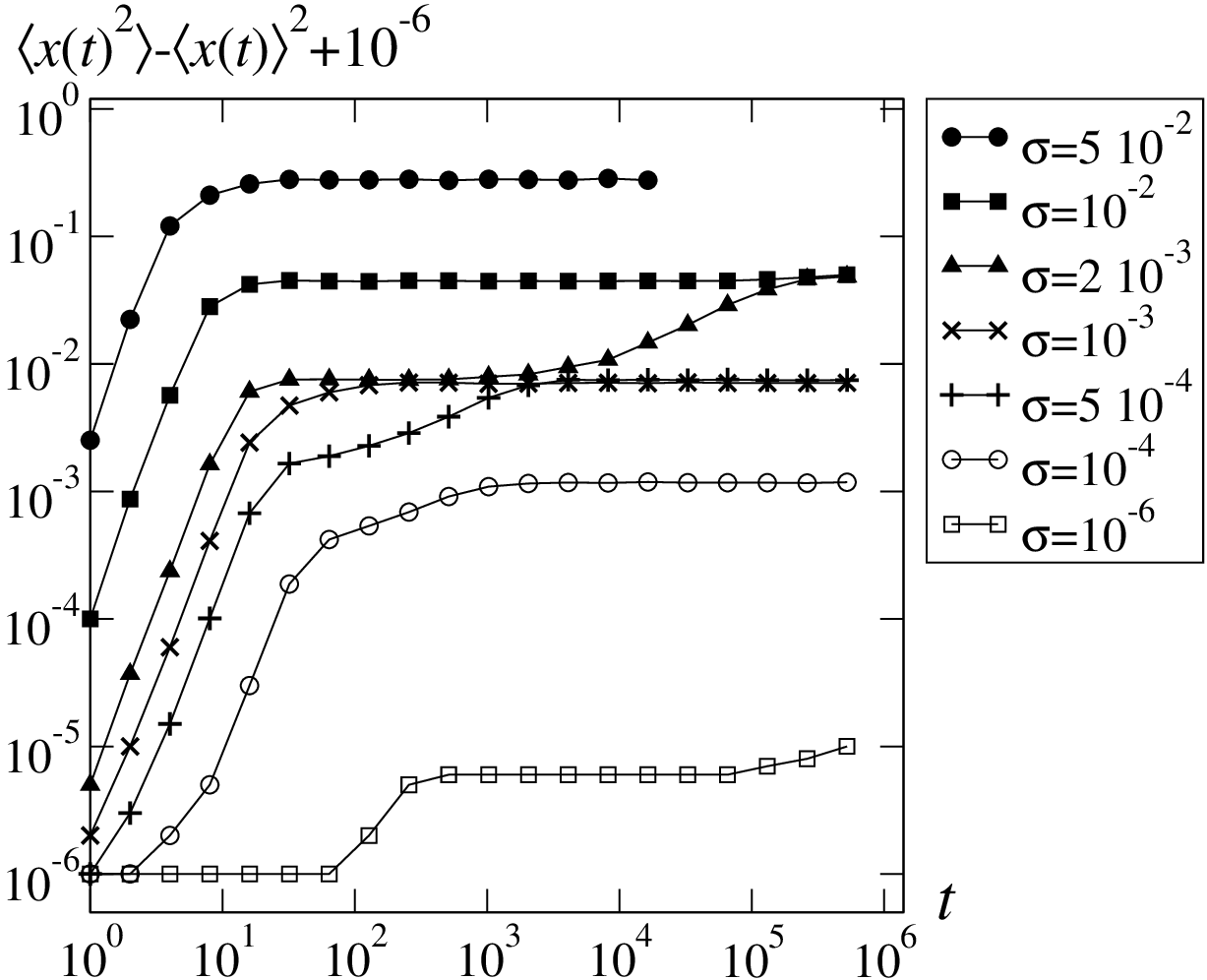}
\end{center}
\caption{\small 
Plateaux structure for the logistic map with
additive noise. Time evolution (using $t=2^n-1$) of the variance
of an ensemble of $M=10^4$ trajectories all starting at $x_0=0$,
for different values the noise strength $\sigma$. 
In order to display data with different orders of magnitude,
the $y$-axis has been plotted in a logarithmic scale with a
shift of $10^6$.
}
\label{fig_logistic_2}
\end{figure}

In order to display the many-plateaux structure of time
evolution in this dynamical model we consider next an
ensemble of $M=10^{3}$ independent copies of the map
(\ref{eq_logistic}) at the edge of chaos $\mu =\mu _{c}$.
The initial positions are $x_{0}=0\;\forall M$ and we
measure the variance 
$\langle x_{t}^{2}\rangle -\langle x_{t}\rangle ^{2}$ (which
plays a role analogous to the {\it temperature} in
the statistical-mechanical description of Hamiltonian systems) 
of the distribution of positions at time
$t=2^{n}-1$. Fig. \ref{fig_logistic_1}(b) shows a numerical
corroboration of previous results \cite{robledo_01}, for
different values of the noise intensity $\sigma $. While in
the first plateau the variance is zero, as the dynamics of
all $M$ trajectories is confined to the attractor, at
the crossover time $t_c\sim \sigma ^{r-1}$ the
ensemble of orbits starts spreading and the variance
increases until a second plateau is reached where it
saturates as a consequence of the confinement induced by the
unstable repeller regions. As it can be well appreciated in
Fig. \ref{fig_logistic_2}, 
%(where data for different $\sigma$
%are represented togheter in a shifted logarithmic $y$-axis), 
for long enough times it is possible to see
another increment of the variance as the iterates are able
to reach other bands of the attractor. The whole process
ends when all bands become occupied and the variance is
consequently of order $1$. 

A two-step process of relaxation is known to be one of the
main dynamical properties displayed by supercooled liquids
close to glass formation \cite {debenedetti_01}. Two other
characteristic features of glassy dynamics are the so-called
Adam-Gibbs formula and the scaling property known as aging.
The first is a relation between the relaxation time
$t_c$ (e.g. the viscosity or the inverse of the
difussivity) and the entropy $S_{conf}$ associated to the
number of possible molecular minimum energy configurations
in the fluid \cite{debenedetti_01}. The second is the loss
of time translation invariance, named aging
\cite{bouchaud_01}, that is due to the fact that properties
of glasses depend on the procedure by which they are
obtained. The time decrease of relaxation functions and
correlations exhibit a scaling dependence on the ratio
$t/t_{w}$ where $t_{w}$ is a waiting time.  As the
counterpart to the Adam-Gibbs formula, it has recently been
shown \cite{robledo_01} that the logistic map at $\mu _{c}$
for $\sigma \neq 0$ presents a relationship between the
plateau duration $t_c$, and the entropy $S_{conf}$
for the state that comprises the largest number of (iterate
positions) bands allowed by noise intensity. This entropy is
obtained from the probability of chaotic band occupancy at
position $x$ \cite {robledo_01}. Also, the trajectories at
$\mu _{c}$ in the limit $\sigma \rightarrow 0$ have been
shown \cite{robledo_01} to obey a scaling property
characteristic of aging, of the form
$x_{t+t_{w}}=g^{(t_{w})}(0)\exp _{q}(-\lambda
_{q}t/t_{w})$ where $g^{(t_{w})}(0)$ is the $t_{w}$-times
composition of the Feigenbaum fixed-point map function
$g(0)$.

\section{QSS in coupled symplectic maps}
\label{section_cons_maps}
Symplectic maps are important tools in the study of
Hamiltonian systems.  If we consider a {\it time-independent}
Hamiltonian system with $n$ degrees of freedom
($2n$-dimensional Gibbs $\Gamma$-space), a
$(2n-2)$-dimensional symplectic map is the 
result of taking a Poincar\'e section over the constant energy
hypersurface \cite{ott_01}. 
The recurrence time is discrete and the map is
useful in displaying the stationary, recurrent properties of
the original Hamiltonian system.  Interestingly enough, a
\mbox{$2n$-dimensional} symplectic map is also  
the result of
a Poincar\'e section in the phase space of a {\it time-dependent} 
Hamiltonian system with $n$ degrees of freedom ($2n+1$
dimensions), with a periodic dependence on time \cite{ott_01}.  
As a
consequence of the symplectic structure, the Lyapunov
spectra in the $d$-dimensional phase space of the map ($d$
being an even number) is characterized by $d/2$ pairs of
Lyapunov coefficients, where each element of the pair is the 
negative of the other.

\begin{figure}
\begin{center}
\includegraphics[width=14cm,angle=0]{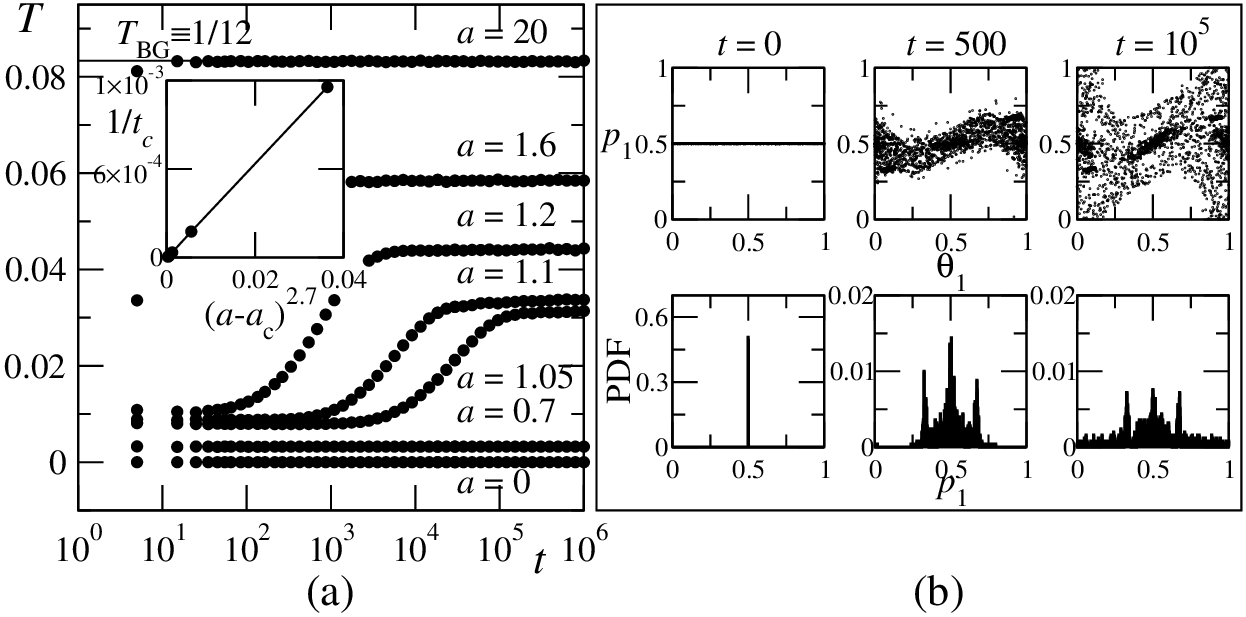}
\end{center}
\caption{\small QSS in the standard map (i.e., $N=1$). (a) $T(t)$ for
typical values of $a$.  We start with ``water bag'' initial
conditions ($M=2500$ points in $0\leq\theta_1\leq 1$,
$p_1=0.5\pm 5\;10^{-4}$).  Inset: Inverse crossover
time $t_c$ (inflection point between the QSS and the BG
regimes with the time axis plotted in logarithmic scale)
vs. $1/(a_1-a_c)^{2.7}$.  (b) Time evolution of the
ensemble in (a) for $a_1=1.1$ (first row) and PDF of its
angular momentum (second row).  $t=0$: Initial conditions;
$t=t_1=500$: The ensemble is mostly restricted by cantori;
$t=t_2=10^5$: The ensemble is confined by KAM-tori.  See
\cite{baldovin_01} for further details.  }
\label{fig_standard_2d}
\end{figure}
A prototypical $2$-dimensional symplectic map, that displays
the KAM mechanism for the transition from regularity to
chaos, is the Chirikov-Taylor {\it standard map} (see, e.g.,
\cite{ott_01} and references therein).  Since fundamental dynamical processes, like
for example Arnold diffusion, occur only for $d>2$, it is
interesting to consider more generally the case of $N$
{\it symplectic} and {\it globally} coupled standard maps ($d=2N$), described, for
example, by the equations:
\begin{eqnarray}
\begin{array}{rclr}
\theta_i(t+1)& =& p_i(t+1) + \theta_i(t) +
\frac{b}{N-1}\sum\limits_{
\substack{
%\begin{subarray}{1}
m=1\\m\neq i
%\end{subarray}
}
}^N p_m(t+1)& \;\;\;({\rm
mod}\;1), \\ p_i(t+1)& =& p_i(t) +
\frac{a}{2\pi}\sin[2\pi\theta_i(t)]& \;\;\;({\rm mod}\;1),
\end{array}
\label{eq_standard}
\end{eqnarray}
where $t=1,2,...$, $i=1,2,...,N$, $b\in{\mathbb R}$ is the
coupling constant (in the following we will fix $b=2$), 
and $a\in{\mathbb R}$ is a parameter that
we will use to control chaoticity;  $\theta_i$ may be
regarded as an angular variable and $p_i$ as an angular
momentum (defined on a compact set though). 
Notice that in order to describe a system with a phase
space of finite size we are considering, as usual, the map on the
torus (${\rm mod}\;1$).  This map is typically chaotic for
large values of $|a|$, while for small $|a|\neq 0$ 
a complex phase-space structure develops of
regular regions surrounded by a stochastic web.
The stochastic web is connected for
$d>2$ (i.e., $N>1$) and disconnected for the marginal case
$d=2$ (i.e., $N=1$) 
(see, e.g., \cite{zaslavsky_01}). 
Intentionally we introduced a global coupling among the
$N$ standard maps: to obtain
time evolution properties similar to those of the
long-ranged interaction 
many-body Hamiltonian that we shall discuss below.

\begin{figure}
\begin{center}
\includegraphics[width=14cm,angle=0]{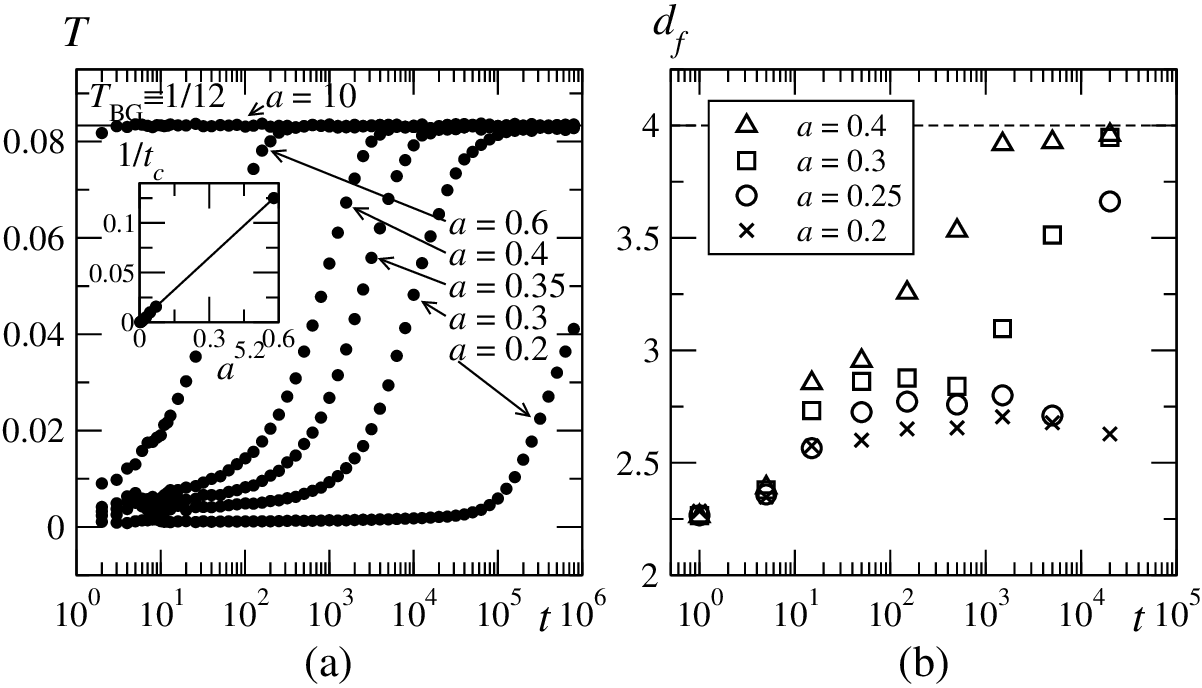}
\end{center}
\caption{\small QSS in two coupled standard maps (i.e., $N=2$).  (a)
$T(t)$ for $b=2$ and typical values of $a$.  We start with
``water bag'' initial conditions ($M=1296$ points with
$0\leq\theta_1,\theta_2\leq 1$, and $p_1,p_2=0.25\pm
5\;10^{-3}$).  Inset: Inverse crossover time $t_c$
vs. $1/a^{5.2}$.  (b) Time evolution of the fractal
dimension of a single initial ensemble in the same setup of
(a).  See \cite{baldovin_01} for further details.  }
\label{fig_standard_4d}
\end{figure}
Our coupled maps display a two-plateau structure with
striking analogies with both the one-dimensional map
discussed in Section \ref{section_logistic} 
and with the HMF model that we
analyze in section \ref{section_HMF}. 
As in the previous section, we start
by considering an ensemble of $M$ independent copies of
map (\ref{eq_standard}). 
In trying to deduce statistical properties directly
from the dynamical behavior of a Hamiltonian system -- with
diagonal kinetic matrix and zero average momentum -- it is
common practice to identify the temperature with the average
squared momentum per particle (see, e.g., \cite{livi_01}). 
Because we
study here situations with nonzero "bulk" motion, a natural
measure of temperature is given by the variance of the total
angular momentum,
\begin{equation}
T(t)\equiv \frac{1}{N}\sum_{i=1}^{N} \left(\langle
p_i^2(t)\rangle-\langle p_i(t)\rangle^2\right),
\label{temperature_standard}
\end{equation}
where $\langle\rangle$ means ensemble average.  As the
classical microcanonical ensemble is based on the equal
a priori probability postulate, we will call {\it BG
temperature} the temperature associated to the uniform
distribution in phase space:
\begin{equation}
T_{BG}\equiv \frac{1}{N}\sum_{i=1}^{N}\left[ \int_0^1
dp_i\;p_i^2-\left(\int_0^1dp_i\;p_i\right)^2 \right]
=1/12\simeq 0.083\;\;\;(\forall N).
\end{equation}
Of course, if the system is sufficiently chaotic or if the
initial ensemble is sufficiently close to equilibrium
(i.e., to the uniform distribution) the temperature
(\ref{temperature_standard}) rapidly relaxes to $T_{BG}$ as
a function of the iteration time.  
Nevertheless, if the phase space presents complex structures 
{\it and} the
initial ensemble is very far-from-equilibrium, partial
barriers (see, e.g.,\cite{mackay_01}) may confine the
ensemble in some limited region for quite long times, before
allowing for a larger occupation of phase space.
Consequently, it appears a plateau with $T=T_{QSS}\neq
T_{BG}$.

\begin{figure}
\begin{center}
\includegraphics[width=14cm,angle=0]{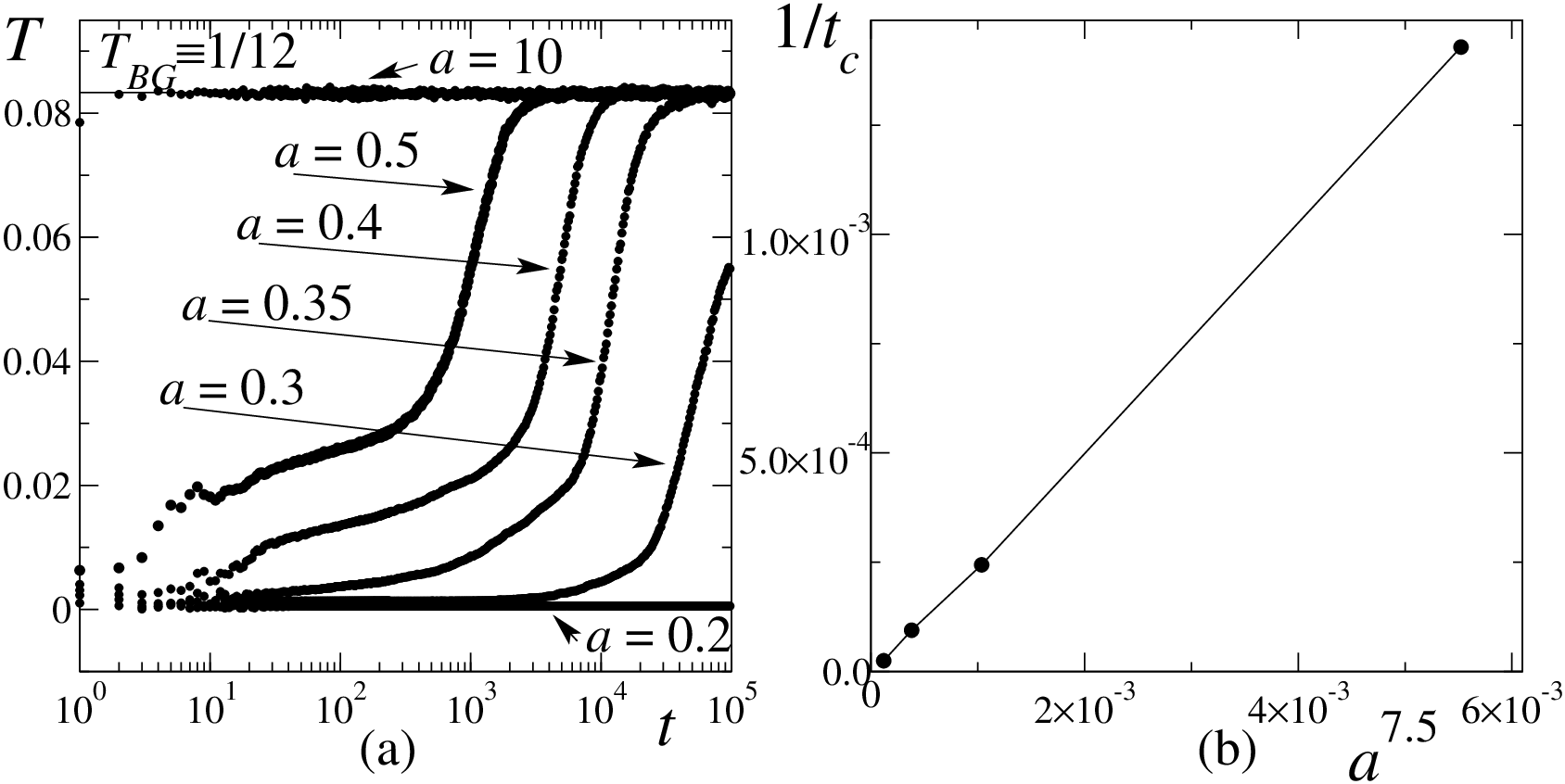}
\end{center}
\caption{\small QSS in $N=100$ coupled standard maps.  (a)
$T(t)$ for $b=2$ and typical values of $a$.  We start with
``water bag'' initial conditions ($M=500$ points with
$0\leq\theta_i\leq 1$ and $p_i=0.25\pm 5\;10^{-3}$ $\forall
i=1,2,...N$).  (b) Inverse crossover time $t_c$
vs. $1/a^{7.5}$.  }
\label{fig_standard_200d}
\end{figure}
In close analogy with the case of section
\ref{section_HMF}, for our out-of-equilibrium initial
conditions we set all momenta $p_i$ randomly distributed
inside a properly chosen small interval of phase space (the
so-called ``water bag'' initial conditions), while the
angles $\theta_i$ are distributed at random inside the whole
interval $[0,1]$.  Fig. \ref{fig_standard_2d}(a) displays
the case of a single standard map ($N=1$). For this marginal
case the formation of total barriers for $a\leq
a_c=0.971635406...$ prevent the formation of a second
plateau below this critical point. We also notice that
unless $a$ is large enough, the temperature associated to
the second plateau is smaller than $T_{BG}$.  However, the
$2$-dimensional case is very useful 
for visualizing the mechanism for plateau formation
(see Fig. \ref{fig_standard_2d}(b)).  
Fig. \ref{fig_standard_4d}(a) indicates that in $d=4$ the
topology of phase space differs from $d=2$. 
Now total barriers for diffusive processes
do not exist, and it is possible to have 
out-of-equilibrium initial data that leads to a dynamics
with a $T_{QQS}$ temperature two-plateau of long duration before
a crossover to $T_{BG}$. 
Interestingly enough, during the first plateau
the geometry of the ensemble is characterized by a
nontrivial fractal dimension ($d_f\simeq 2.7<4=d$) that at
equilibrium crosses over to the dimension of the whole phase
space (see Fig. \ref{fig_standard_4d}(b)).  
As shown by the new results presented in
Fig. \ref{fig_standard_200d}(a) for the 
$200$-dimensional phase space obtained by coupling $N=100$
standard maps, the familiar two-plateau structure persists
for large values of $N$. Preliminary results suggest that
even for large values of $N$ a first plateau 
develops with $d_f/d<1$. This fact strongly reinforces the
picture that there are nonzero-measure sets of initial
conditions such that the system is maintained dynamically in
a subset of the whole phase space, and only later 
occupies the entire allowed space, and adopts a BG-like
statistics. This behavior is the more pronounced the closer
to zero is the largest Lyapunov exponent.

\begin{figure}
\begin{center}
\includegraphics[width=14cm,angle=0]{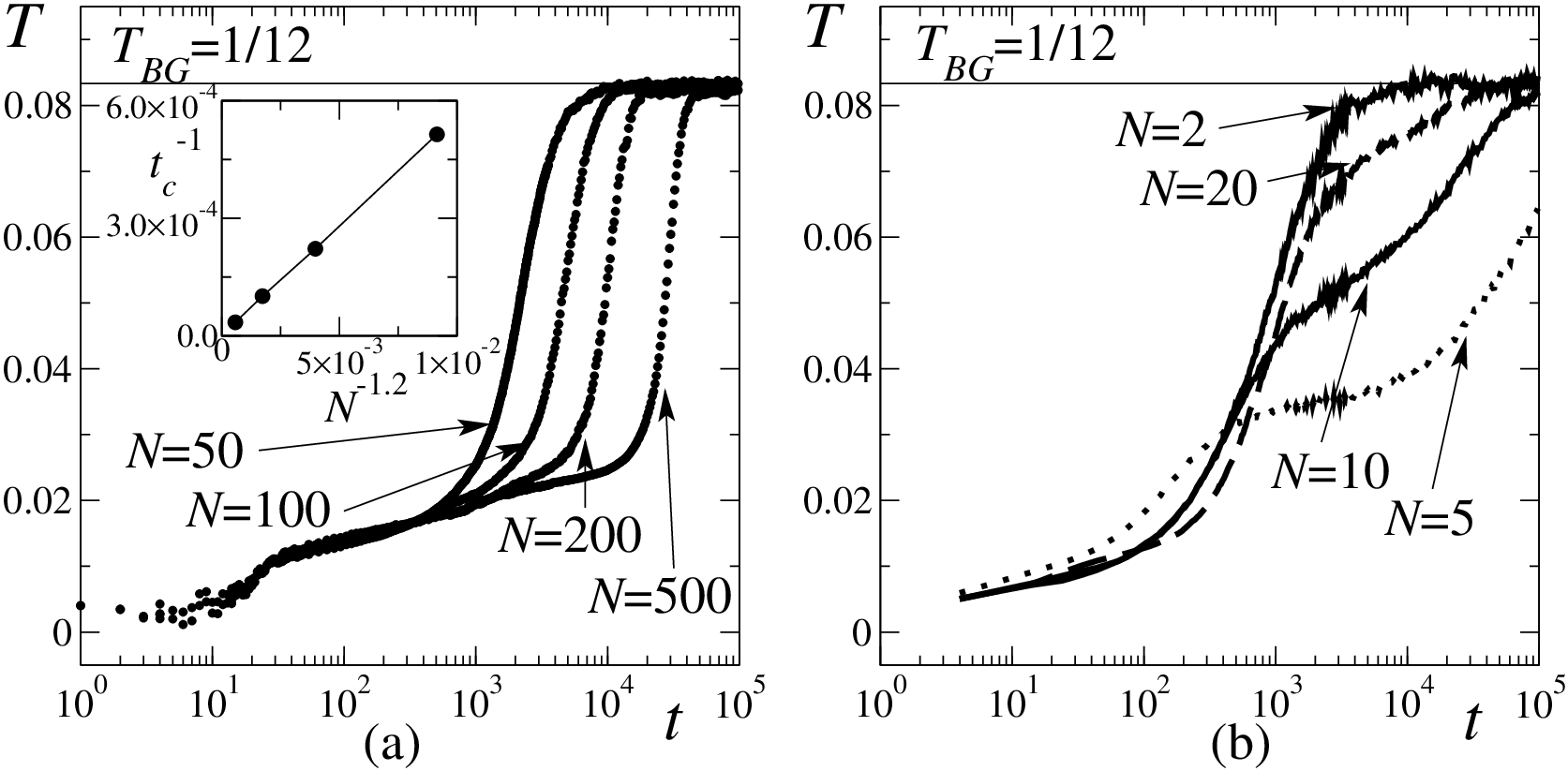}
\end{center}
\caption{\small 
Scaling of the crossover time with the number of coupled
standard maps.
Initial data are defined as in Fig.  \ref{fig_standard_200d}.
(a) $T(t)$ for $a=0.4$, $b=2$ and large values of $N$.  
Inset: Inverse crossover time $t_c$
vs. $1/N^{1.2}$.
(b) $T(t)$ for $a=0.4$, $b=2$ and small values of $N$.  
}
\label{fig_standard_N}
\end{figure}
It is important to stress
that in all these cases the crossover time to BG statistics
diverges as $a\to 0$ (or as $a\to a_c$ for $d=2$), as it is
shown in 
Fig. \ref{fig_standard_200d}(b) and in the inset of
Fig. \ref{fig_standard_4d}(a) and \ref{fig_standard_2d}(a).
The consequence of this is that if the $t\to\infty$ limit is taken
after the $a\to 0$ (or $a\to a_c$ for $d=2$) limit, the
temperature associated to these ensembles is forever $T_{QSS}\neq
T_{BG}$.
The same feature is revealed if, fixing $a$, we analyze the
scaling of the crossover time for a large $N$
(see Fig. \ref{fig_standard_N}(a) where we have taken $a=0.4$).
Moreover, the exponent in the scaling relation 
$t_c\sim N^{1.2}$ is very close to what have been found for
the QSS of the HMF model: $t_c\sim N$ \cite{latora_01}.
It is also interesting to notice that, for the case we are
discussing, this scaling is effective only for large value
of $N$ (say, $N>20$). 
For small $N$ a more intricate, richer behavior appear
(see Fig. \ref{fig_standard_N}(b)). 
In particular, for $N=5$ 
we observe for the first time in the case of globally
coupled symplectic maps evidence of many-plateau structures
(like those in section 2).

\section{QSS in the Hamiltonian Mean Field model}
\label{section_HMF}
Along the previous lines of thought we can draw our
attention now to a
more complex dynamical scenario, that is, Hamiltonian
many-body dynamics. 
In analogy with the globally-coupled symplectic maps of the
previous section, we will here 
focus on {\it long-range} Hamiltonians. Long-ranged
interacting systems constitute nowadays an important subject
in many areas of physics, e.g. astrophysics, hydrodynamics,
nuclear physics, Bose-Einstein condensates, atomic clusters,
plasma physics, and several others \cite{dauxois_01}. If
interaction terms decay as $r^{-\alpha}$ and $d$ is the
dimension of the system, we say that a classical 
system has long-range interactions when $\alpha/d>1$.
Long range interactions induce long range correlations, and,
as a consequence of this, it has been found that these
systems show, in a variety of situations, sensible
deviations from the BG equilibrium behavior. 

In this section we will overview some known aspects and
present as well new results 
for the HMF model \cite{antoni_01}. 
This model has been largely considered in the 
literature,  
as it has been found to display a surprisingly rich behavior 
(see, e.g., \cite{dauxois_02} for a review). 
Here we are particularly interested in the appearance, under
specific initial conditions (see below), of QSSs that after a
certain amount of time cross over to the BG equilibrium. 
During the QSSs a variety of anomalous behaviors
has been detected.
Among them, 
vanishing Lyapunov spectrum,
L\'evy walks and anomalous diffusion, 
anomalous distribution of the momenta, 
non-commutability of the infinite time limit with 
the thermodynamical limit
(see, e.g., \cite{tsallis_02} and references therein), 
and, more recently, aging \cite{montemurro_01,pluchino_01}  
and glassy behavior \cite{pluchino_02}.   
As some connections have already been established with
nonextensive statistical mechanics
\cite{tsallis_02,montemurro_01,pluchino_01}, 
these QSSs are 
candidates for a situation
that is correctly described by this formalism.

The HMF Hamiltonian has the following form: 
\begin{equation}
H = K+V=\sum_{i=1} ^{N} \frac{p_{i}^2}{2} + 
\frac{1}{2N} \sum_{i,j=1} ^{N} 
\left[1-cos(\theta_{i}-\theta_{j})\right],
\label{H}
\end{equation}
where $\theta_i\in[0,2\pi)$ is the $i$th angle and
$p_i\in{\mathbb R}$ is the
conjugate variable  
representing the angular momentum (with unit inertial
momenta).  It is an inertial version of the XY
ferromagnetic spin model where the interaction terms 
couple globally all the spins. 
Because of the presence of the kinetic part in the
Hamiltonian, this model (in opposition with the usual
Ising-like spin models) is naturally provided of a
dynamics. 
Also note that it is common use (but not strictly
necessary \cite{anteneodo_01}) to divide the potential term
by $N$ in order to make the Hamiltonian (formally)
extensive \cite{kac_01}.

This system can be analytically solved within the BG
canonical formalism, which predicts a second-order phase
transition from a low-energy ferromagnetic phase with
magnetization $m\equiv|{\mathbf m}|\neq 0$, where 
${\mathbf m}\equiv \sum_{i=1}^N{\mathbf m}_i/N$ 
(with ${\mathbf
m}_i=(\cos\theta_i,\sin\theta_i)$), to a high-energy phase
with all spins uniformly distributed on the interval
$[0,2\pi)$ (and consequently $m=0$). 
The critical point is at specific energy
\mbox{$u=u_c\equiv 0.75$}, and the complete equilibrium
caloric curve $T_{BG}(u)$ can be exactly
obtained \cite{dauxois_02}.
On the other hand, it is possible to numerically integrate 
the Hamilton equations at fixed total energy,
therefore allowing an analysis in the microcanonical
setup. 
Results show nonequivalence between the two
approaches for a range of energy densities below the critical
point and for a certain class of initial conditions (which includes the 
so-called ``water bag'' initial conditions, characterized by
a random distribution of the angular momenta around zero).  
Numerical
simulations show in fact a first regime where the system
stays
in a non-equilibrium metastable
state. It is possible to
distinguish this QSS by observing the evolution of
the dynamical temperature, here defined in the
standard way $T(t)\equiv 2K(t)/N$ since the total angular
momentum is zero. 
After a short transient the system evolves into a
first stage with a temperature $T_{QSS}<T_{BG}$ and
it is only at longer times that it relaxes to the predicted
equilibrium value. 
Very similarly to the case discussed in the previous
section, the duration of the QSS diverges linearly as $N$ goes to
infinity \cite{latora_01}. 

\begin{figure}
\begin{center}
\includegraphics[width=10cm,angle=0]{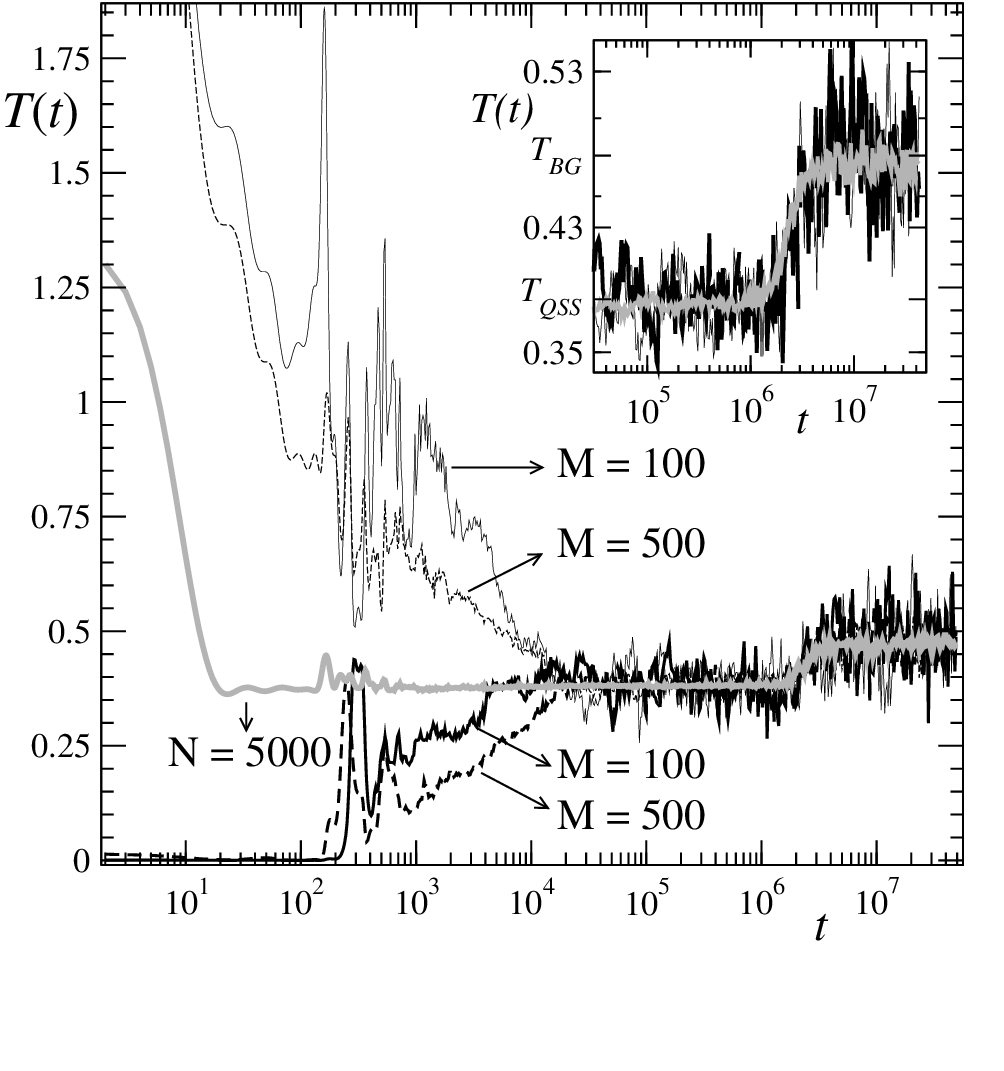}
\end{center}
\caption{\small Temperature evolution of an isolated
$N=5000$ spin system (Eq. (\ref{H})) for $u=0.69$ in grey
line. We start with ``water bag'' initial conditions for the
momenta (an ``almost uniform'' distribution inside the
range $-2\lesssim p_i\lesssim 2$: 
See \cite{moyano_01} for details)  
and magnetization $m=1$ ($\theta_i=0\;\forall i$).
\emph{Hot} (starting from above) and \emph{cold}
(starting from below) smaller subsystems are shown, both
with $M=100$ (solid thin line) and $500$ (dashed thick
line). Inset: Magnification of the crossover between
$T_{QSS} \simeq 0.38$ and $T_{BG} \simeq 0.467$.}
\label{fig_HMF_T}
\end{figure}
Inside this framework, we present numerical results using a
textbook construction (due to Gibbs in fact) of the canonical ensemble as a
subsystem of the microcanonical one. 
It is a purely dynamical approach, 
in contrast to the analytical derivation
mentioned above. 
We consider in fact a subset of $M$ spins of the
total $N$ spins in the isolated (microcanonical)
system, and study the temperature evolution of this subset,
particularly during the metastable state. Temperature is defined
as $T_M(t)\equiv 2 K_M(t)/M$, where $K_M$ is the kinetic
energy of the $M$ spins.  Different criteria may be used in
order to choose the $M$ subsystem. We show two examples,
namely selecting the highest and the lowest 
temperature subsystem. 
Result are illustrated  in Fig. \ref{fig_HMF_T} for $u=0.69$
\cite{moyano_01}: 
Two curves coming from above and below, respectively. In both
cases $T_M$ is seen to relax after some time to $T_N$,
\emph{within the metastable state.} Afterwards, both
temperatures evolve sharing the same value and perform
together the crossover to the predicted $T_{BG}$. This
result is repeated for two system sizes, $M=100$ (solid
line) and $M=500$ (dashed line), yielding the same
behavior. 
It is important to stress that in the case of the
hot subsystems, during the relaxation to $T_{QSS}$
the temperature value $T_{BG}$ is crossed with no signs of 
relaxation to it. 
Simulations are made using the $4$th
order symplectic Neri-Yoshida integrator \cite{yoshida_01} with
energy conservation $\Delta U /U \simeq 10^{-4}$, and
``water bag'' initial condition.
This result gives a first step into the study of thermal
meta-equilibrium of two systems, i.e., the zeroth principle of
thermodynamics for the QSS and, furthermore, 
allows for the possibility
of a generalized canonical treatment of the metastable
state, which constitutes in fact an ongoing study \cite{baldovin_03}.

\begin{figure}
\begin{center}
\includegraphics[width=14cm,angle=0]{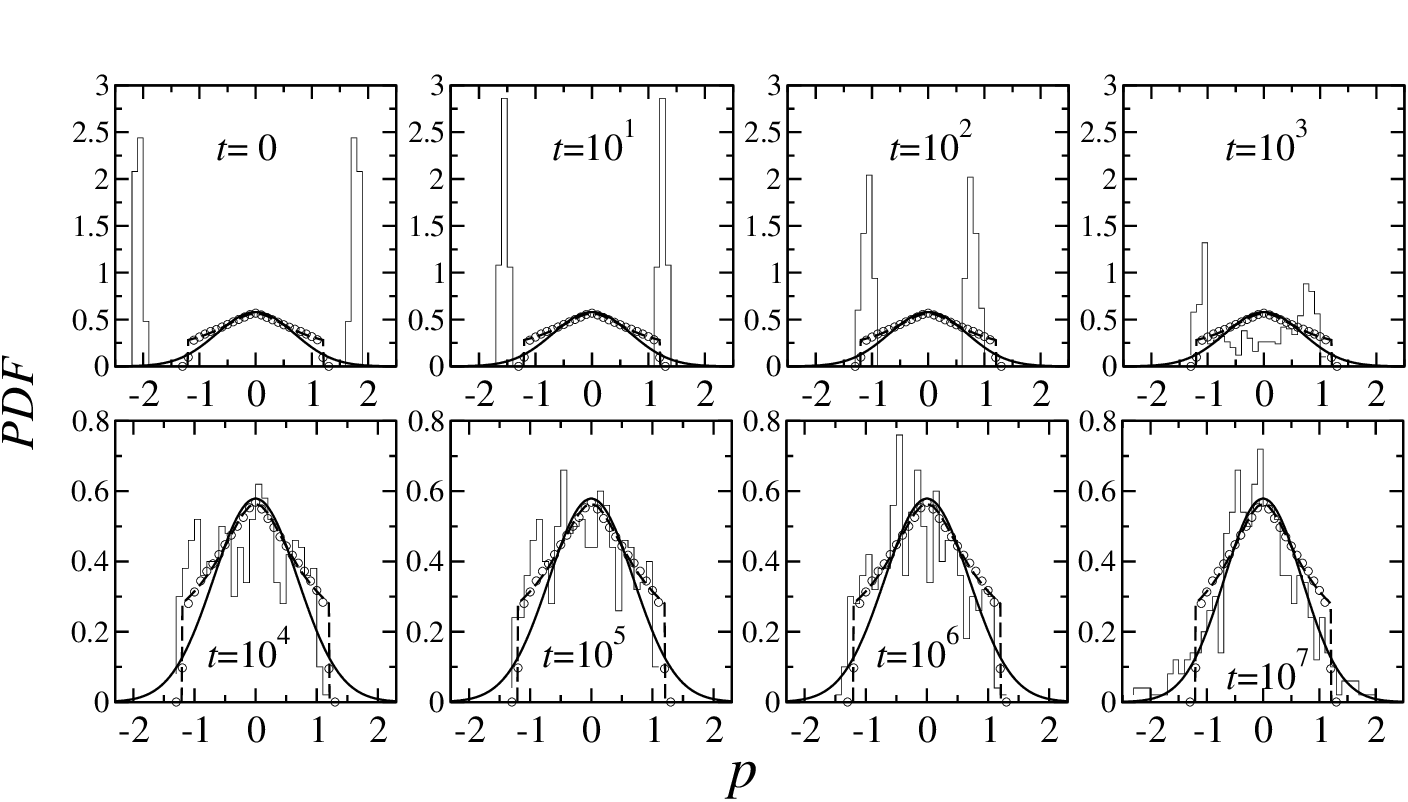}
\end{center}
\caption{\small Snapshots of the evolution of PDF
for momenta in the $\mu$-space,
starting with a ``hot water bag'' (corresponding to
dashed thin curve in Fig. \ref{fig_HMF_T}). 
The staired histogram shows 
the instantaneous PDF at time $t=0$ and $t=10^k$
(with $k=1,2,...7$), for $M=500$ spins inside an $N=5000$ spin
system. In empty circles, the average of $10^3$ realizations
for $N=10^5$ spins (all averages made during the $QSS$
plateau). In dashed line, the $q$-exponential fitting curve
with $T=T_{QSS}=0.38$ and $q=3.7$. Finally, the curve in
solid line shows the 
analytical Gaussian PDF. These last three curves are the
same in every frame, and are plotted for reference.}
\label{fig_HMF_PDF}
\end{figure}

We also present a new analysis of how the PDF of the momenta
of the $M$-subset evolves with time.
Results are given in
Fig.\ref{fig_HMF_PDF}. 
We observe that the PDF for the $M$-subset essentially
coincides with that of the total $N$-system.
After a transient of order $t\sim 10^3$, the PDF stabilizes
into an non-Maxwellian distribution that lasts up to the
crossover time $t\sim 10^6$. During this whole metastable stage
the PDF is well fitted by a $q$-Maxwellian with a
cut-off, with temperature $T=T_{QSS}$ and $q\simeq 3.7$.
The value of $q$ that we find is different from that
calculated in \cite{latora_01}. 
This may be due to the fact that we are using slightly
different (more ordered) initial conditions or to finite-size effects.
Our present conjecture is that not only the value
of $q$ can change when the thermodynamic limit is adequately 
approached (first $N \to \infty$, then $M \to \infty$, and
finally $t \to \infty$), but also the location of the
cut-off (in momenta) may go further away, towards infinity. 
After the crossover time the PDF leaves the QSS anomalous
distribution and relaxes to the equilibrium quasi-Maxwellian with
$T=T_{BG}$ (strictly Maxwellian, with no cutoff at all, only for $N \rightarrow \infty$).

\section{Summary and discussion}
\label{section_discussion}

Let us summarize our results. We have reviewed and analyzed three quite
different dynamical systems: First, the logistic map at the edge of chaos
with additive noise, secondly a system composed of globally coupled standard
maps, and finally the HMF model. In all three cases a common dynamical feature
is found: The occurrence, for certain classes of initial conditions, of
metastable states that after a certain time evolve into the stable equilibrium
state. 
These metastable or quasistationary states appear to be intimately associated to
a partial, non-trivial, occupation of phase space, typically (multi)fractal
(scale-free and perhaps network-like). This feature retains the system in a
non-ergodic state (while held in reserve from entering into all regions of the
phase space available to the system).
A detailed description for this situation can be given for the simpler case of
the logistic map with additive noise at the onset of chaos, i.e., the interplay
between a multifractal attractor and a multifractal repeller. This
circumstance can also be recognized in the more complex system of two coupled
standard maps, since our calculations clearly show the occurrence of quasistationary
states and their crossover to final standard chaotic behavior. As we have seen
too, the characteristic two-plateau feature is maintained for the case of 
hundreds of coupled standard maps, and so, we may propose the hypothesis that a
fractal occupation holds even for a many-dimensional phase space. The
similarities observed between globally coupled symplectic maps and the HMF
model indicate that the same state of affairs possibly holds in the case of
the QSS detected in the HMF model, a system for which a direct analysis of the
$\Gamma$-space is much harder.

The importance of these results is underscored by the fact that the QSSs do
become permanent when conventional limits are taken in a specific order, e.g.
the thermodynamic limit before the infinite time limit in the HMF model. There
is a nonuniform convergence feature in the dynamical evolution that leads to
an atypical stationary state. Non-commutability of limits occurs too in the
case of the maps. 
For the coupled standard maps
this is appreciated when the thermodynamic limit is replaced
(for fixed $a$)
by the limit of infinite number of coupled maps, or when the
chaoticity parameter $a$ tends (for fixed $N$) towards a specific value
($a=0$ for $d>2$, and $ a_c$ if $d = 2$); 
while in the case of the logistic map it is the limit of vanishing
noise that plays an analogous role. 
Several other common features are found for the
QSSs occurring in these different systems. 
For instance, {\it in all of them the whole Lyapunov spectrum
vanishes}, which seems to pinpoint the cause of the observed
anomalies. 
Also, aging and other
glassy properties are 
both present in the logistic map with additive noise and in the HMF model.
These common features suggest a deeper, basic, connection among these systems.
As we have seen, the usual BG statistical theory appears to be inadequate
in explaining the metastable states, and that a generalization of the usual BG
statistics is required. In this respect, the nonextensive theory [6] comes
out as the strongest candidate to meet this test.

\section*{Acknowledgments}
We acknowledge C. Anteneodo and A. Rapisarda for useful discussions and
comments. 
FB, LGM and CT have benefitted from partial support by FAPERJ,
CNPq and PRONEX (Brazilian agencies). 
APM was partially supported by SECYT-UNC (Argentinean agency).
AR was partially supported by CONACyT and
DGAPA-UNAM (Mexican agencies).

\end{document}